\documentstyle[12pt,]{article}

\begin{document}

\begin{center}

{\Large \bf \sc{Planckian
Scattering Beyond the Eikonal Approximation in the Functional Approach}}\\

{\large {\bf \small {\sc{Nguyen Suan Han}}} and {\bf \small
{\sc{Nguyen Nhu Xuan }}}$\footnote {Permanent address: Department
of Theoretical Physics, Vietnam National University, P.O. Box 600,
BoHo, Hanoi 10000, Vietnam; e-mail: han@phys-hu.edu.vn \bf{
Submitted to European Physical Journal C.}} $ } \vspace*{0.3cm}

\end{center}

\centerline{Institute of
Theoretical Physics, Chinese Academy of Sciences,}
\centerline{P.O. Box 2735, Beijing 100080,  China }
\centerline{12th March 2002}


\begin{abstract}

\small {In the framework of functional integration   the
non-leading terms to leading eikonal behavior of the Planckian-energy
scattering amplitude  are calculated by the
straight-line path approximation. We show that the allowance for the
first-order correction terms leads to the appearance of
retardation effect. The singular character of the correction terms
at short distances is also noted, and they may be lead ultimately
to the appearance of non-eikonal contributions to the scattering
amplitudes.\\

\noindent
PACS 11.80-Relativistic scattering theory.\\
PACS 04.60-Quantum gravity.}

\end{abstract}

\section{Introduction}

    The asymptotical behavior of scattering amplitude in high energy
is one of the center problems of elementary particle physics. The
standard  method of the quantum field theory is expected  that the
calculations based on the perturbation theory  are suitable when
the energy of individual particles is not rather high and the
effective coupling constant is not large. When the energy is
increased the effective coupling constant also increases so that
the corrections calculated by the perturbation theory play a
crucial role. Gravitational scattering at Planckian-energy $
\sqrt{s}=2E \geq M_{PL}$, where $s$ is the square of the centre of
mass energy,  $ M_{PL}$ is the Planck mass, and small angle are
characterized by the effective coupling constant $ \alpha_{G}=
\frac{Gs}{\hbar}\geq 1 $ which makes any simple perturbative
expansion unwarranted. Comparison of the results of the different
approaches $[1-8]$ proposed for this problem has shown that they all
coincide in the leading order approximation which has a
semiclassical effective metric interpretation, while most of them
fail in providing the non-leading terms under which new classical
and quantum effects are hiding.$[2,3]$\\

\noindent
The aim of the present paper is to continue the
determination of the non-leading terms to the Planckian-energy scattering
by functional approach proposed for constructing a scattering
amplitude in our previous works $[9,10]$. Using the straight-line
path approximation we have shown that in the limit of
asymptotically high $ s \gg M_{PL}^{2} \gg t $,  at the fixed momentum
transfers $ t$ the lowest order eikonal expansion of the exact
two-particle Green function on the mass shell gives the leading
behavior  of the Planckian-energy scattering amplitude, which
agrees with the results found by all others $[1-8]$.  The main
advantage of the proposed approach is the possibility of
performing calculations in the compact form and obtaining the sum of
the considered diagrams immediately in a closed form.\\

\noindent
The outline of the paper is as follows. In the second section
using the example of scalar model $ L_{int} =g{\varphi}^2\phi$
which allows to make an exposition  most clarity and descriptive,
and also less tedious calculations, by means of the functional
integration, we briefly demonstrate the conclusion of the leading
behavior $[9-17]$  and explain the important steps in  calculating
the non-leading terms to the high-energy scattering amplitude
$[15]$.  This section can be divided into  three parts. In the first
the quantum Green function of two-particles is obtained in the
form of the functional integral. In the second part  by a transition
to the mass shell of the external two particle Green function we
obtain a closed representation for the two-particle scattering
amplitude which is also expressed in the form of the functional
integrals. In the last of this section the straight-line path
approximation and its generalization  are discussed for
calculating the non-leading terms to high-energy scattering
amplitudes. Based on the exact expression of the single-particle
Green function in gravitational field $g_{\mu\nu}(x)$ obtained in
Ref.$[9]$ the results discussed in the second section will be
generalized in the third section to the case of scalar "nucleons"
of the field $\varphi (x) $ interacting with a gravitational
field. Finally in the fourth section, we draw our conclusions.

\section{ Corrections to the Eikonal Equations in Scalar Model }

In the construction of a scattering amplitude we use a reduction formula
which relates an element of the $S$ matrix to the vacuum expectation
of the chronological product of the field operators. For the
two-particle amplitude, this formula has the form
$$
i(2\pi)^4 \delta^4 (p_1+p_2-q_1-q_2)T(p_1,p_2;q_1,q_2)= $$
$$=i^4\int \prod_{k=1}^{2} dx_k dy_k \overrightarrow{K}_{x_1}^{m}
\overrightarrow{K}_{x_2}^{m}
\langle0|T(\varphi(x_1)\varphi(x_2)\varphi(y_1)\varphi(y_2))|0\rangle
\overleftarrow{K}_{y_1}^{m}\overleftarrow{K}_{y_2}^{m},
\eqno(2.1)$$\\
\noindent
where  $p_1$, $p_2$ and $q_1$, $ q_2$ are the moments of the
particles of the field $\varphi(x)$ before and after scattering
respectively.\\

\noindent
Ignoring the vacuum polarization effects the two-nucleon Green
function on the right-hand side of Eq.$(2.1)$  can be represented in
the form
$$
G(x_1,x_2;y_1,y_2)=\langle0|T(\varphi(x_1)\varphi(x_2)\varphi(y_1)\varphi(y_2))|0\rangle=$$
$$
=\exp\Biggl \{\frac{i}{2}\int D\frac{\delta^2}{\delta\phi^2}\Biggl
\}\Biggl[G(x_1,y_1|\phi)G(x_2,y_2|\phi)+
G(x_1,y_2|\phi)G(x_2,y_1|\phi)\Biggl]\Bigg|_{\phi=0},\eqno(2.2)
$$
where
$$
\exp\Biggl \{\frac{i}{2}\int D\frac{\delta^2}{\delta\phi^2}\Biggl
\}=\exp\Biggl\{ \frac{i}{2}\int d^4z_1 d^4z_2
D(z_1-z_2)\frac{\delta^2}
{\delta\phi(z_1)\delta\phi(z_2)}\Biggl\}, \eqno(2.3)$$
\noindent
and $G(x,y|\phi)$ is the Green function of the nucleon
$\varphi(x)$ in a given external field $\phi(x)$.\\
\noindent
The nucleon Green function $G(x,y|\phi)$ satisfies the equation
$$
[\Box +m^2 -g\phi(x)]G(x,y|\phi)=\delta^4(x-y), \eqno(2.4)
$$
\noindent
whose formal solution can be written in the form of  Feynman path
integral
$$
G(x,y|\phi)=i\int_{0}^{\infty}e^{-im^2 \tau}d\tau \int[\delta^4\nu]_{0}^{\tau}\\
 \exp \Biggl \{\ ig \int dz J (z)\phi (z)\Biggl \}
\delta^4\Biggl(x-y+2\int_{0}^{\tau}\nu(\eta)d\eta\Biggl),
\eqno(2.5)$$
\noindent
where $ J(z)$ is the classical current of the nucleon \footnote{In
the scalar model $J(z)$ describes the spatial density of nucleon
moving on a classical trajectory. However, in this case we call
$J(z)$ a current.}
$$
J(z)=\int_0^\tau d\eta \delta^4 \Biggl( z-x+2\int_0^{\tau}
\nu(\xi) d\xi  \Biggl ), \eqno(2.6)
$$
$[{\delta}^{4}\nu_{i}]_{\tau_1}^{\tau^2}$ is a volume element of
the functional space  of the four-dimensional function $
\nu(\eta)$ defined on the interval
$ \tau_1 \leq \eta \leq \tau_2$,\\

$$ [{\delta}^{4}\nu_{i}]_{\tau_1}^{\tau^2}=
\frac{ {\delta}^{4}\nu_{i} \exp [-i\int_{\tau_1}^{\tau^2}
{\nu}_{\mu}^{2}(\eta) \prod_{\eta} d^{4} {\eta}} {\int
{\delta}^{4}\nu_{i} \exp [-i\int_{\tau_1}^{\tau^2}
{\nu}_{\mu}^{2}(\eta) \prod_{\eta} d^{4} {\eta } }. $$

Substituting Eq.$(2.5)$ into Eq.$(2.2)$ and performing the variational
differentiation with respect to $\phi$, we find that the Fourier
transform of the two-nucleon Green function
$$
G(p_1,p_2;q_1,q_2)= \int \prod_{i=1}^{2} \Biggl (d^4x_i d^4y_i
e^{i(p_i x_i-q_iy_i)} \Biggl )G(x_1,x_2;y_1,y_2)     \eqno(2.7)
$$
is given by the following expression

$$ G(p_1,p_2|q_1,q_2)
=i^2 \prod_{i=1}^{2} \Biggl(\int_0^\infty d\tau_i
e^{i\tau_i(p_i^2-m^2)}\int[ \delta ^4 \nu_i ]_{0}^{\tau_i} \int
dx_i e^{ix_i(p_i-q_i)} \Biggl)$$
$$\times \exp\Biggl[-\frac{ig^2}{2}\int D(J_1+J_2)^2 \Biggl]
+ (p_1\leftrightarrow p_2), \eqno(2.8)
$$
where we have introduced the abbreviated natation
$$
\int J_i D J_k=\int\int dz_1 dz_2 J_i(z_1) D(z_1-z_2) J_k (z_2).
\eqno(2.9)
$$

Expanding the expression $(2.8)$ with respect to the coupling
constant $g^2$ and taking the functional integrals with respect
to $\nu_i $, which reduce to simple Gaussian quadratures if a Fourier
transformation is made, we obtain  the well-known series of
perturbation theory for $ G(p_1,p_2|q_1,q_2)$.\\

  The elastic-scattering amplitude is related to the two-nucleon
Green function as

$$i(2\pi)^4 \delta^4(p_1+p_2-q_1-q_2)T(p_1,p_2|q_1,q_2)^{scalar}$$
$$=\lim_{p_i^2,q_i^2 \rightarrow m^2
}\Biggl(\prod_{i=1,2}(p_i^2-m^2)(q_i-m^2)\Biggl)G(p_1,p_2|q_1,q_2)
+ (p_1\leftrightarrow p_2). \eqno(2.10)
$$
\noindent
Substituting Eq.$(2.5)$ to Eq.$(2.2)$ and making a number of substitutions
of the functional  variables $[9]$, we obtain a closed expression
for  the two-nucleon scattering amplitude in the form of
functional integrals

$$ T(p_1,p_2;q_1,q_2)^{scalar}=\frac{g^2}{(2\pi)^4}\int d^{4}x
e^{i(p_1-q_1)x} D(x)$$
$$\times\Biggl (  \prod_{i=1}^{2}\int
[{\delta}^{4} {\nu_i}]_{-\infty}^{\infty} \exp \Bigr \lbrace
{i\frac{g^2}{2}\sum_{i=1,2} \int \Bigr (  J_{i} D J_{i} -i\delta_i
m^2 \Bigl) } \Bigl \rbrace\Biggl)$$
$$\times \exp \int_{0}^{1}d\lambda
 \exp\Biggl(ig^2\lambda \int J_1DJ_2\Biggl)
 +(p_1\leftrightarrow p_2),\eqno(2.11) $$

\noindent
where the quantity $ J_i(z, p_i, q_i | \nu_i )$ is a conserving
transition current given by

$$ J_i(z, p_i,q_i | \nu_i )=\int_{-\infty}^{\infty}d{\xi}
 \delta \Biggl( z-x_i - a_i(\xi)
+2\int_{0}^{\xi} {\nu}_i (\eta) d{\eta} \Biggl),\eqno(2.12) $$

$$
a_{1,2}(\xi)=p_{1,2}\theta(\xi)+q_{1,2}\theta(-\xi). \eqno(2.13)
$$\\

  The scattering  amplitude $ (2.11)$ is interpreted as the residue
of the two-particle Green function $ (2.8) $ at the poles
corresponding to the nucleon ends. A factor of the type $ \exp
\Bigr ( -\frac {i{\kappa}^2}{2} \sum_{i=1,2}\int J_{i} D J_{i}
\Bigl)$ of Eq.$(2.11) $ takes into account the radiative
corrections to the scattered nucleons, while $ \exp
\Bigr(i{\kappa}^2\lambda e^{ikx}\int J_1DJ_2 \Bigl) $ describes
virtual-meson exchange among them. The integral with respect to $
d\lambda $ ensures the subtraction of the contribution of the freely
propagating particles from the matrix element. The functional
variables $ \nu_1(\eta)$ and $\nu_2(\eta)$ formally introduced for
obtaining the solution of the Green function describe the
deviation of a particle trajectory from the straight-line paths.
The functional with respect to $ [{\delta}^{4}{\nu}_i]$ $(i=1,2)$
corresponds to the summation over all possible trajectories of the
colliding particles. From the consideration of the integrals over
$ \xi_{1} $ and $ \xi_{2} $ for $ \exp \Bigr ( -\frac
{i{\kappa}^2}{2} \sum_{i=1,2}\int J_{i} D J_{i}\Bigl)$ it is seen
that the radiative correction  result in divergent expressions of
the type $ \delta_{i} m^{2}\times ( A \rightarrow \infty )$. To
regularize them, it is necessary to renormalize the mass, that is,
to separate from  $ \exp \Bigr ( -\frac {i{\kappa}^2}{2}
\sum_{i=1,2}\int J_{i} D J_{i}\Bigl)$ the terms $ \delta_i
m^{2}\times ( A \rightarrow \infty ), (i=1,2) $, after which we go
over in Eq.$(2.11)$ to the observed mass $ {m_i}_{R}^2=
{m_i}_{0}^2 + \delta_i m^2 $.  These problems have been discussed in
detail in previous works $[9,10,12,18]$, therefore we shall
hereafter drop the radiation corrections terms
$\exp \Biggl(i\frac{g^2}{2}\sum_{i=1,2}  \int \Bigl[J_{i} D J_{i} -i\delta_i
m^2 \Bigl]\Biggl)$ as these contributions in our model can be
factorized as a factor $R(t)$ that depends only on the
square of moment transfer. A similar factorization of
the contributions of radiative corrections in quantum
electrodynamics has also been obtained $[19]$.\\

Ignoring the radiation corrections, the elastic-scattering
amplitude of two scalar nucleons $ (2.11)$ can be presented in the
following form

$$
T(p_1,p_2|q_1,q_2)^{scalar}=\frac{ig^2}{(2\pi)^4}\int d^4 x
e^{-ix(p_1-q_1)}D(x) \int_{0}^{\lambda} d\lambda S_{\lambda} +
(p_1\leftrightarrow p_2), \eqno(2.14)
$$
where
$$
S_{\lambda}=  \int \prod_{i=1}^{2}
[\delta^4\nu_i]_{-\infty}^{\infty}\exp\{ig^2\lambda \Pi[\nu]\} ;
\Pi[\nu]= \int J_1DJ_2,  \eqno(2.15)$$
\noindent
and the quantity $J_i(k,p_i,q_i|\nu_i)$ is a conserving transition
given by

$$
J_i(k,p_i,q_i|\nu_i)=\int_{-\infty}^{\infty}d\xi \exp\Biggl (
2ik[a_i(\xi)+\int_{0}^{\xi}\nu_i(\eta)d\eta] \Biggl).\eqno(2.16)
$$
\noindent
Note that the expression $(2.12)$ defines the scalar density of a
classical point particle moving along the curvilinear path $
x_i(s) $, which depends on the proper time $ s=2m\xi $ and
satisfies the equation

$$
mdx_i(s)/ds= p_i \theta(\xi) +q_i \theta(-\xi)+\nu_i(\xi)
\eqno(2.17)
$$
\noindent
subject to the condition $ x_i(0)=x_i, i=1,2 $. For this reason,
the representation $(2.11)$ of the scattering amplitude can be
regarded as an functional sum over all possible nucleon paths in
the scattering process.\\

However, functional integrals $(2.14)$ cannot be integrated
exactly and an approximate method must be developed. The simplest
possibility is to eliminate $ \nu_i(\xi) $ from the argument of  $
J_i (k,p_i,q_i|\nu_i)$ function, i.e., we set $\nu_i(\xi) =0 $ in
Eq.$(2.16)$ for transition current, and obtain

$$J_i( k, p_i,q_i |\nu_i)=
 \Biggr[\frac{1}{2p_ik +i\epsilon} -\frac{1}{2q_i k-i\epsilon}
\Biggl] ,\eqno (2.18)
$$
which corresponds to the classical current of a nucleon  moving with
momentum $p $  for $\xi> 0$ and momentum  $q $ for $\xi<0$. \\

Note however that the approximation $\nu =0 $ is certainly false
for proper time $ s $ of the particle near rezo, when the
classical trajectory of the particle changes direction. In the
language of Feynman diagrams, this corresponds to neglecting the
quadratic dependence on $ k_i $ in the nucleon propagators, i.e.,

$$ \Bigl [m^2- (p- \sum_{i=1} ^{n} k_{i} )^2 \Bigl]^{-1} \rightarrow
  \Bigl [ 2p\sum_{i=1} ^{n} k_{i} \Bigl ]^{-1},\eqno(2.19) $$
\noindent
which can lead to the appearance of divergences of integrals
with respect to $ d^4k $ at the upper limit. As is well known,
this approximation $(2.19)$ can be used to study the infrared
asymptotic behavior in quantum electrodynamics $[13,20,
21]$. However, it has not been proved in the region of
high energies  $[11-13]$.\\

Therefore, we shall use an approximate method of calculating integrals with respect to
$\nu_i(\xi)$ which enables one to retain the quadratic dependence of
the nucleon propagators on the momenta $ k_i$. This method is based
on the following expansion formula $[13,14,22]$
$$
\overline{\exp{\Bigl(g^2\Pi[\nu]\Bigl)}}= \int [\delta^4 \nu ]\exp
\Bigl(g^2\Pi[\nu]\Bigl) =\exp\Bigl(g^2\overline{\Pi[\nu]}\Bigl)
\Biggl[1+\sum_{n=2}^{\infty}\frac{(g^2)^n}{n!}
\overline{(\Pi-\overline{\Pi})^n}\Biggl],\eqno(2.20)
$$
where $ \overline{\Pi[\nu]}=\int [\delta^4 \nu ]|\Pi[\nu]$.

Applying the modified expansion formula $(2.20)$ exposed in detailed
in Ref.$[15]$ to our case, we consider the leading term $(n=0)$
and the following correction term $(n=1)$. When $ n=0 $ the leading
term has the form

$$
S_{\lambda}^{(n=0)scalar}=\overline{\exp{\Bigl(i\lambda
g^2\Pi[\nu]\Bigl)}}=\int [\delta^4 \nu]\exp (i \lambda
g^2\Pi[\nu])\approx \exp (i \lambda g^2\int [\delta^4
\nu]\Pi[\nu]),\eqno(2.21)$$

\noindent
where

$$
\left.\overline{\Pi[\nu]}\right|_{\nu =0}=\frac{1}{(2\pi)^4} \int
d^4 k D(k)\exp(-ikx) $$
$$ \times \int_{-\infty}^{\infty}d\xi d\tau
\exp \Biggl (2ik \Bigl [\frac{\xi a_1(\xi)}{\sqrt{s}}-\frac{\tau
a_2(\tau)}{\sqrt{s}}\Bigl]\Biggl ) \times\exp
\left[i\frac{k^2}{\sqrt{s}}(|\xi|+|\tau|)\right]. \eqno(2.22)
$$
\noindent
In Eq.$(2.22)$, we have made the change of variables
$\xi,\tau\rightarrow \xi/\sqrt{s},\tau/\sqrt{\tau}$.
When $n=1$ the correction term has the following form
$$
S_{\lambda}^{(n=1)scalar}= \exp (i\lambda g^2 \overline{\Pi[\nu]})
 \exp \left.\left[ 1+\frac{i\lambda^2 g^4}{4} \Biggr(
\int d\eta \sum_{i=1,2}
 \Biggr(\frac{\delta\overline{\Pi[\nu]}}{\delta\nu_i(\eta)}\Biggr)^2 \Biggr )
\right]\right|_{\nu=0}. \eqno(2.23) $$\\
\noindent
Using Eq.$(2.22)$ we have
$$
\frac{i\lambda^2 g^4}{4}\int d\eta
\left[\Biggr(\frac{\delta\overline{\Pi[\nu]}}{\delta\nu_1(\eta)}\Biggr)^2
+\Biggr(\frac{\delta\overline{\Pi[\nu]}}{\delta\nu_2
(\eta)}\Biggr)^2 \right]= \frac{i\lambda^2 g^4}{(2\pi)^8}\int
d^k_1 d^4 k_2 e^{-ix(k_1+k_2)}D(k_1) D(k_2)(k_1k_2)$$
$$
\times\int_{-\infty}^{\infty} d\xi_1 d\tau_1 d\xi_2 d\tau_2
\exp\left \{2ik_1 \left[
\xi_1\frac{a_1(\xi_1)}{\sqrt{s}}-\tau_1\frac{a_2(\tau_1)}{\sqrt{s}}
\right]
\left[i\frac{k_1^2}{\sqrt{s}}(|\xi_1|+|\tau_1|)\right]\right\}$$
$$\times
\exp\left \{2ik_2 \left[
\xi_2\frac{a_1(\xi_2)}{\sqrt{s}}-\tau_1\frac{a_2(\tau_2)}{\sqrt{s}}
\right]
\left[i\frac{k_2^2}{\sqrt{s}}(|\xi_2|+|\tau_2|)\right]\right\}
\frac{1}{\sqrt{s}}[\Phi(\xi_1,\xi_2)+\Phi(\tau_1,\tau_2)],
\eqno(2.24)
$$
where
$$
\Phi(\xi_1,\xi_2)=\vartheta(\xi_1,\xi_2)
[|\xi_1|\vartheta(|\xi_2|-|\xi_1|)+|\xi_2|
\vartheta(|\xi_1|-|\xi_2|)],
$$
$$
\Phi(\tau_1,\tau_2)=\vartheta(\tau_1,\tau_2)
[|\tau_1|\vartheta(|\tau_2|-|\tau_1|)+|\tau_2|
\vartheta(|\tau_1|-|\tau_2|)]. \eqno(2.25)
$$
In this approximation the nucleon propagator functions in Eqs
$(2.21)-(2.25)$ do not contain the terms of type $k_ik_j $, where
$k_i$ and $k_j$ belong to different mesons interacting with
nucleons. This means that in nucleon propagators we can neglect
the terms of  form $ \sum_{i\neq j}k_ik_j $ compared with $
2p\sum_{i} k_i $, i.e., we can make the substitution

$$ \Bigl [m^2- (p- \sum_{i=1} ^{n} k_{i} )^2 \Bigl ]^{-1} \rightarrow
 \Bigl [ 2p\sum_{i=1} ^{n} k_{i} - \sum_{i=1}^{n} k_{i}^{2}\Bigl ]^{-1}.\eqno(2.26) $$

\noindent
This approximation $k_ik_j=0 $ which is called the straight-line
path approximation  corresponds to the approximate calculation of the
Feynman  path integrals $[9-17]$ in Eq.$(2.11)$ and $(2.14)
$ in accordance with the rule $(2.26)$.  The formulation of
the straight-line path approximation made it possible to put
forward a clear physical conception. In accordance with
which high-energy particles move along Feynman paths that
are most nearly rectilinear. \\

   The validity  of the given approximation of  Eq.$(2.26)$
in the region of high energies  $ s $ for given momentum transfers
$t $ can be studied within the framework of perturbation theory.
In particular, one can show that neglecting   the terms $ k_ik_j=0
$ the denominators of the nucleon propagator functions in the case
of ordinary ladder diagrams obtained by iteration of the
single-meson exchange diagram does not effect the asymptotic
behavior at high energies, which, when mesons are exchanged, has
form ${ln}s/s^{n-1}$. The validity of this approximation $(2.26) $
has also been proved for larger class diagrams with interacting
meson lines $[16]$. In addition, it should be noted that the
eikonal approximation in the potential scattering also reduces to
a modification of the propagator (which is nonrelativistic in
this case),  a modification determined $[25]$ by Eqs $(2.19)$ and
$(2.26)$.\\

We shall seek the asymptotic behavior of the functional integral $
S_{\lambda}$ at large $s=(p_1+p_2)^2$ and fixed momentum transfers
$ t=(p_1-q_1)^2 $. For this , we go over to the center -of-mass
system and take the $ z $ axis along the moment of the incident
particles. Then

$$
p_{1,2}=\Biggl \{\frac{\sqrt{s}}{2},0,0,
\pm\frac{\sqrt{s-4m^2}}{2}\Biggl \};
$$
$$
q_{1,2}=\Biggl \{\frac{\sqrt{s}}{2},\pm
\bigtriangleup_{\perp}\sqrt{1+\frac{t}{s-4m^2}}
\pm\frac{\sqrt{s-4m^2}}{2}
\Biggl(1+\frac{2t}{s-4m^2}\Biggl)\Biggl \}. \eqno(2.27)
$$
$$
\triangle_{\bot}^2=-t.
$$
\noindent
And, substituting Eq.$(2.27)$ into Eq.$(2.14)$, we obtain

$$
a_{1,2}(\xi)=\frac{1}{\sqrt{s}}\Biggl[p_{1,2}\theta(\xi)+q_{1,2}\theta(-\xi)\Biggl]=
\frac{1}{2}\Biggl[ \theta(\xi)+\theta(-\xi)\Biggl]$$
$$\pm\Biggl( \frac{\Delta_{\bot}}{\sqrt{s}}
\sqrt{1+\frac{t}{s-4m^2}}\Biggl)\theta(-\xi)
\pm \frac{\sqrt{s-4m^2}}{\sqrt{s}}\Biggl(1+\frac{t}{s-4m^2}\Biggl).
\eqno(2.28)
$$
In the limit $ s \rightarrow\infty$ for fixed $t$ and keeping the terms
to order $O(\frac{1}{s})$,we found
$$
\frac{a_1(\xi)}{\sqrt{s}}\approx\frac{1}{2}n^{+}+
\frac{\triangle_{\bot}}{\sqrt{s}}\vartheta(-\xi)+O\Bigl(\frac{1}{s}\Bigl
),
$$
$$
\frac{a_2(\xi)}{\sqrt{s}}\approx\frac{1}{2}n^{-}-
\frac{\triangle_{\bot}}{\sqrt{s}}\vartheta(-\xi)+O\Bigl(\frac{1}{s}\Bigl),
$$
$$
n^{\pm}=\{1,0,0,\pm1 \}. \eqno(2.29)
$$

We now find the asymptotic behavior of the expressions $(2.22)$
and $(2.24)$ as $ s\rightarrow\infty $ and fixed $ t $.
Using Eq.$(2.29)$, we obtain an asymptotic expression for
Eqs $(2.22)$ and $(2.24)$. Namely

$$
\overline{\Pi[\nu]}=\frac{1}{(2\pi)^6 s}\int d^4 k e^{-ikx}D(k)
\int_{-\infty}^{\infty}d\xi d\tau e^{i( k_{-}\xi-
k_{+}\tau)}\times$$
$$
\times\Biggl\{1-2i\frac{k_{\bot}\triangle_{\bot}}{\sqrt{s}}[\xi\vartheta(-\xi)
+\tau\vartheta(-\tau)]+\frac{ik^2}{\sqrt{s}(|\xi|+|\tau|)}\Biggl
\}$$
$$\approx -\frac{1}{8\pi^2s}\int\frac{d^2 k_{\perp}}{k_{\perp}^2+\mu^2}
e^{ik_{\perp}x_{\perp}}+$$

$$
+\frac{i\triangle_{\bot}}{s\sqrt{s}8\pi^2}
[x_{+}\vartheta(-x_{+})-x_{-}\vartheta(x_{-})] \int d^2 k_{\perp}
e^{ik_{\perp}x_{\perp}} \frac{k_{\perp}}{k_{\perp}^2+\mu^2}+$$
$$
+\frac{i}{16\pi^2 s\sqrt{s}}(|x_{+}|+|x_{-}|)
\int\frac{d^2k_{\perp}}{k_{\perp}^2+\mu^2} e^{ik_{\perp}x_{\perp}}$$
$$
=- \frac{1}{4\pi s}K_0(\mu|x_{\perp}|)- \frac{\mu}{4\pi
s\sqrt{s}}\frac{\bigtriangleup_{\perp}x_{\perp}}{|x_{\perp}|}
[x_{+}\vartheta(-x_{+})-x_{-}\vartheta(x_{-})]
K_1(\mu|x_{\perp}|)-$$
$$
- \frac{i\mu^2}{8\pi
s\sqrt{s}}(|x_{+}|+|x_{-}|)K_0(\mu|x_{\perp}|)
,\eqno(2.30)$$

\noindent
where $ x_{\pm}=x_0\pm x_z $, the light cone
coordinates,  $k_{\pm}^{(i)}=k_{0}^{(i)} \pm k_{z}^{(i)}$, $i=1,2$
and $\mu$ is the mass of the changed particle, which must  be
introduced as an infrared regulator. The final expression is

$$
\frac{i\lambda^2 g^4}{4}\int d\eta
\left[\Biggr(\frac{\delta\overline{\Pi[\nu]}}{\delta\nu_1(\eta)}\Biggr)^2
+\Biggr(\frac{\delta\overline{\Pi[\nu]}}{\delta\nu_2
(\eta)}\Biggr)^2 \right]\approx $$
$$
\approx -\frac{i\lambda^2 g^4}{(2\pi)^8 s^2 \sqrt{s}} \int d^4 k_1
d^4 k_2 D(k_1) D(k_2) \exp [-ix(k_1+k_2)](k_1k_2)$$
$$\times\int_{-\infty}^{\infty} d\xi_1 d\tau_1
e^{i(k_{-}^{(1)}\xi_1 -k_{+}^{(1)}\tau_1)}
\int_{-\infty}^{\infty}d\xi_2 d\tau_2
e^{ i(k_{-}^{(2)}\xi_2 -k_{+}^{(2)}\tau_2)}
\Biggl[\Phi(\xi_1,\xi_2)+\Phi(\tau_1,\tau_2)\Biggl]
$$
$$=-  \frac{i\lambda^2 g^4 \mu^2}{32\pi^2 s^2
\sqrt{s}}(|x_{+}|+|x_{-}|) K_1^2(\mu|x_{\perp}|), \eqno(2.31)
$$
\noindent
here we have assumed $|x_{\perp}|\neq 0$, which ensures that all the
integrals converge. The functions $ K_0(\mu|x_{\perp}|)$ and
$K_1(\mu|x_{\perp}|)$ are MacDonald functions of the zeroth and first
orders and are determined by the expressions
$$
K_0(\mu|x_{\perp}|)=\frac{1}{2\pi}\int d^2 k_{\perp}
 \frac{\exp(ik_{\perp}x_{\perp})}{k_{\perp}^2+\mu^2},$$

$$K_1(\mu|x_{\perp}|)=-\frac{\partial K_0(\mu|x_{\perp}|)}
{\partial(\mu|x_{\perp}|)}. \eqno(2.32)
$$

We now substitute Eqs $(2.30)$ and $(2.31)$ into Eq.$(2.24)$ and obtain
for the correction term $ S_{\lambda}^{(n=1)}$ the desired
expression

$$
S_{\lambda}^{(n=1)}\approx \exp \Bigl [-\frac{ig^2 \lambda}{4\pi
s}K_0(\mu|x_{\perp}|)\Bigl ] \Biggl\{1-\frac{ig^2 \lambda\mu}{4\pi
s\sqrt{s}} \frac{\triangle_{\bot}x_{\bot}}{|x_{\bot}|}
[x_{+}\vartheta(-x_{+}) -x_{-}\vartheta(x_{-})]
K_1(\mu|x_{\perp}|) $$
$$+ \frac{g^2 \lambda\mu^2}{8\pi s\sqrt{s}}
(|x_{+}|+|x_{-}|)K_0(\mu|x_{\perp}|)
-\frac{ig^4\lambda^2\mu^2}{32\pi^2 s^2 \sqrt{s}}
(|x_{+}|+|x_{-}|)K_1^2(\mu|x_{\perp}|)\Biggl \}. \eqno(2.33)
$$
In this expression $(2.33)$ the factor in the front of the braces
corresponds to the leading eikonal behavior of the scattering
amplitude, while the terms in the braces determine the correction
of relative magnitude $1/\sqrt{s}$.\\

     As is well known from the investigation of the
scattering amplitude in the Feynman diagrammatic technique, the
high-energy asymptotic behavior can contain only logarithms and
integral powers of $ s $. A similar effect is observed here, since
integration of the expression $(2.33)$ for $ S_\lambda $ in
accordance with Eq. $(2.14)$ leads to the vanishing of the
coefficients for half-integral powers of $s$. Nevertheless,
allowance for the terms that contain the half-integral powers of
$s$ is needed for the calculations of the next corrections in the
scattering amplitude. It is interesting to note the appearance in
the correction terms of a dependence on $ x_0 $ and $ x_z $
$( x_{\pm}=x_0\pm x_z) $, i.e.,
the appearance of the so-called retardation effects, which are
absent in the  principal asymptotic term.\\

    Making similar calculations, we can show that all the following
terms of the expansion $(2.20)$ decrease  sufficiently rapidly compared
with those we have written down. However, it must be emphasized
that this by no means proves the validity of the eikonal
representation for the scattering amplitude in the given framework.
The coefficient functions in the asymptotic expansion, which
are expressed in terms of MacDonald functions, are singular at
short distances and this singularity becomes stronger with
increasing rate as decrease of the corresponding terms at  large
$ s $. Therefore, integration of $ S_\lambda $ in accordance with
Eq.$(2.14)$ in the determination of the scattering amplitude may
lead to the appearance of terms that violate the eikonal series in
the higher order in $ g^2 $. The possible appearance of such terms
in individual orders of perturbation theory in models of  type
$\varphi^3 $ was pointed out in Refs $[23]$, $ [24]$ and $[16] $.
Investigating the structure of the non-eikonal contributions to
the two-nucleon scattering amplitude show  that the sum of
all ladder diagrams of the eighth order in the scalar model contains
terms that are absent in the orthodox eikonal equation and vanish
in the limit $(\mu/m)\rightarrow 0$,  where $\mu $ and $ m $
are meson and nucleon masses. These terms correspond to the contributions
to the effective quasipotential resulting from exchange of
nucleon-antinucleon pairs $[28]$.\\

    To conclude the section we consider the asymptotic behavior of the
elastic scattering amplitude of two scalar nucleons Eq.$(2.14)$
in the ultra-high-energy limit $s\rightarrow \infty $,  $t/s \rightarrow 0$.
In this case the phase function of the leading eikonal behavior
$ \chi(b,s)= -\frac{g^2}{4\pi s}K_0(\mu|x_{\perp}|)$ followed from
Eq.$(2.33)$ does not depend on $ x_{+} $ and $x_{-} $.  Performing the
integration $dx_{+}$, $dx_{-}$ and $d\lambda$ for scattering amplitude
in the center-of-mass (c.m.s) system\footnote {The amplitude $T(s,t)$ is normalized in the c.m.s.
by relation

$$
\frac{d\sigma}{d\Omega}=\frac{|T(s,t)|^2}{64\pi^2 s},
\qquad \sigma_{t}=\frac{1}{2p\sqrt{s}}{Im} T(s, t=0).
$$ } we obtain the following eikonal form

$$
T(s,t)= -2is\int d^2x_{\perp} e^{i\Delta_{\perp}x_{\perp}}
\Bigl(e^{-i\chi(x_{\perp}s)}- 1 \Bigl), \eqno(2.34)
$$
where $ x_{\perp}$ is a two-dimensional vector perpendicular to
the nucleon-collision direction (the impact parameter),
the eikonal phase function $ \chi(x_{\perp}s)$ by scalar
meson exchange decreases with energy
$$
\chi(x_{\perp},s)=\frac{g^2}{4\pi s}K_0(\mu|x_{\perp}|).
\eqno(2.35)
$$
\noindent
 For a similar calculation it has been shown  that the exchange
term $( p_1 \leftrightarrow p_2) $ is one order $(1/s)$ smaller and so
can be dropped in Eq.$(2.33)$. The amplitude is in an eikonal
form. The case of interaction of nucleons with vector mesons, graviton,
can be treated in a similar manner.\\

\section{ Corrections to the Eikonal Equations in Quantum Gravity }

In the framework of standard field theory for the high-energy
scattering the different methods have been developed to investigate the
asymptotic behavior of individual Feynman diagrams and their
subsequent summation.  The calculations of eikonal diagrams in the
case of gravity produce in a similar way as analogous
calculations in QED.  The eikonal captures the leading behavior of
each order in perturbation theory, but the sum of leading terms is
subdominant to the terms neglected by this approximation. The
reliability of the eikonal amplitude for gravity is uncertain. One
approach which has probed the first of these features with some
success is that  based on reggeized string exchange amplitudes
with subsequent reduction to the gravitational eikonal limit
including the leading order corrections $[2,26,27]$.
In this paper we follow a
somewhat different approach based on a representation of the
solutions of the exact equation of the theory in the form of
functional integral.  By this approach we obtain the closed
relativistically invariant crossing symmetry  expressions for
the two-nucleon elastic scattering amplitudes $[9]$, which may be
regarded as sum over all trajectories  of colliding nucleon
and are helpful to investigate the asymptotical behavior of
scattering amplitudes in different kinematics at low to high
energies.\\

  We consider  the scalar nucleons  $\varphi(x) $ interacting with
gravitational field  $ g_{\mu\nu}(x)$ where the interaction
Lagrangian is of the form

$$
L(x)=\frac{\sqrt{-g}}{2}  [g^{\mu\nu}(x) \partial_{\mu} \varphi(x)
\partial_{\nu}\varphi(x) - m^2\varphi^2 (x) ] +L_{grav.}(x) , \eqno(3.1) $$

\noindent
where $ g={det}g_{\mu\nu}(x)=\sqrt {-g} g^{\mu\nu} (x)$. For the
single-particle Green function in the gravitational field
$g^{\mu\nu}(x)$ in the harmonic coordinates defined the condition
$
\partial_{\mu}{\tilde g^{\mu\nu}(x)}=0 $, we have the following
equation

$$    [\tilde g^{\mu\nu}(x)i\partial_{\mu}i\partial_{\nu}
-\sqrt{-g} m^2 ]G(x,y|g^{\mu\nu})=\delta^4(x-y),\eqno(3.2)$$

\noindent
whose solution can be written in the form of a functional integral
$[9]$

$$ G(x,y|g^{\mu\nu})=i\int_{0}^{\infty} d{\tau} e^{-im^2 \tau} $$
$$\times C_{\nu} \int {\delta}^{4}{\nu} \exp \Biggl ( -i\int_{0}^{\tau} d{\xi}
[\tilde g^{\mu\nu} (x,\xi) ]^{-1} {\nu_{\mu}(\xi)}{\nu_{\nu}(\xi)}
- im^{2} \int_{0}^{\tau} [ \sqrt{-g(x_{\xi})} -1]d{\xi} \Biggl)$$
$$\times {\delta}^{4}\Biggl( x-y-2\int_{0}^{\tau} \nu(\eta)
d{\eta} \Biggl). \eqno (3.3) $$
\noindent
Equation $(3.3) $ is the exactly closed expression for the scalar
particle Green function in an arbitrary  external gravitational
field $g^{\mu\nu}(x)$ in the form of functional integrals $[9]$.\\

    In the following we consider the gravitational field in
the linear approximation, i.e., put $ g^{\mu\nu}={\eta}_{\mu\nu} +
\kappa h_{\mu\nu} $,  where $ {\eta}_{\mu\nu} $ is the Minkowski
metric tensor with diagonal $ (1,-1,-1,-1) $.\\
Rewrite Eq.$(3.3)$ in the variables $ h_{\mu\nu}(x) $ and after
dropping out the term with an exponent power higher than the first
$h_{\mu\nu}(x)$ \footnote{ The Lagrangian $(3.1)$ in the linear
approximation to $ h^{\mu\nu}(x)$ has the form  $
L(x)=L_{0,{\varphi}}(x)+L_{0,grav.}(x) +L_{int}(x) $,  where

$$ L_{0}(x)=\frac{1}{2} [\partial^{\mu} \varphi (x)\partial_{\mu} \varphi(x)
-m^{2} {\varphi}^{2}(x) ] ,$$

$$ L_{int}(x)=-\frac{\kappa}{2} h^{\mu\nu}(x)T_{\mu\nu}(x) ,$$

$$ T_{\mu\nu}(x)=\partial_{\mu}\varphi(x) \partial_{\nu}\varphi(x)-
\frac{1}{2}\eta_{\mu\nu} [\partial^{\sigma} \varphi
(x)\partial_{\sigma} \varphi(x) -m^{2} {\varphi}^{2}(x) ] ,$$
\noindent
where $T_{\mu\nu}(x)$ is the energy momentum tensor of the scalar field.
The coupling constant $\kappa $ is related to Newton's constant of
graviton G by $ \kappa^2=16\pi G $. } , we have a Green function
for single-particle Klein-Gordon equation in a linearized
gravitational field

$$ G(x,y|h^{\mu\nu})=i\int_{0}^{\infty} d{\tau} e^{-im^2 \tau}
\int [{\delta}^{4}\nu]_{0}^{\tau} \exp \Biggl( i\kappa \int
J_{\mu\nu}(z) h^{\mu\nu}(z) dz \Biggl){\delta}^{4}
\Biggl(x-y-2\int_{0}^{\tau}\nu(\eta)d{\eta}\Biggl),\eqno(3.4) $$

\noindent
where $ J_{\mu\nu}(z)$ is the current of the nucleon defined by

$$ J_{\mu\nu}(z)=\int_{0}^{\tau_i} d{\xi}
 (\nu_{\mu}(\xi) \nu_{\nu} (\xi))
{\delta}\Biggl (z- x_i+2p_i\xi + 2 \int_{0}^{\xi} {\nu}_i(\eta)
d{\eta} \Biggl).\eqno(3.5) $$

Substituting Eq.$(3.4)$ to Eq.$(2.2)$ and making analogous calculations
done in Ref.$[9]$, for scattering amplitudes we obtain the
following expression

$$ T(p_1,p_2;q_1,q_2)^{tensor}=
{\kappa^2}\int d^{4}x e^{i(p_1-q_1)x} \Delta(x; p_1, p_2; q_1, q_2
) \int_{0}^{1} d\lambda S_{\lambda}+ (p_1\leftrightarrow p_2),
\eqno(3.6) $$

\noindent
where

$$ S_{\lambda}^{tensor}=  \int \prod_{i=1}^{2}
[\delta^4\nu_i]_{-\infty}^{\infty}\exp\{i\kappa^2\lambda
\Pi[\nu]\} , \Pi[\nu]=\int J_1DJ_2\, \eqno(3.7)$$

$$\Delta (x; p_1, p_2; q_1,q_2 )=
\int d^4{k} D^{\mu\nu\rho\sigma}(k)e^{ikx}$$
$$\times [k+p_1+q_1]_{\mu}[k+p_1+q_1]_{\nu}
[-k+p_2+q_2]_{\rho}[-k+p_2+q_2]_{\sigma}. \eqno(3.8) $$

\noindent
The quantity $ J_i^{\mu\nu}(k; p_i,q_i | \nu_i )$ in Eq. $(3.7)$ is a
conserving transition current given by

$$ J_i^{\mu\nu}(k;p_i,q_i|\nu )=4\int_{-\infty}^{\infty}d{\xi}
[a_i(\xi) + {\nu}(\xi) ]^{\mu} [a_i(\xi) +{\nu}(\xi)]^{\nu}\exp
\Biggl( 2ik[\xi_i a_i(\xi) + \int_{0}^{\xi} {\nu}_i (\eta) d{\eta}
]\Biggl),\eqno(3.9) $$

\noindent
and $ D_{\alpha\beta\gamma\delta}(x) $ is the causal Green
function

$$
D_{\alpha\beta\gamma\delta}(x)=
\omega_{\alpha\beta,\gamma\delta}\frac{i}{(2\pi)^4} \int
\frac{e^{ikx}}{k^2 - \mu^{2} + i\epsilon} d^4k ,$$

$$ \omega_{\alpha\beta,\gamma\delta}=(\eta_{\alpha\gamma}\eta_{\beta\delta}
+\eta_{\alpha\delta}\eta_{\beta\gamma}-\eta_{\alpha\beta}\eta_{\gamma\delta}).
  $$
The leading term $(n=0)$ and the following correction term $(n=1)$ in
the case of quantum gravity can be constructed in a similarly way
done in the scalar model,

$$
S_{\lambda}^{(n=0)tensor}=\int [\delta^4 \nu]\exp (i \lambda
g^2\Pi[\nu])\approx \exp (i \lambda \kappa^2\int [\delta^4
\nu]\Pi[\nu]), \eqno(3.10)$$
\noindent
where
$$
 \left.\overline{\Pi[\nu]}\right|_{\nu =0}=
\frac{1}{(2\pi)^4} \int d^4 k e^{-ikx} \int_{-\infty}^{\infty}d\xi
d\tau a_{1}^{\mu}(\xi)a_{1}^{\nu}(\xi)D_{\mu\nu\sigma\varrho}(k)
a_{2}^{\sigma}(\tau)a_{2}^{\varrho}(\tau)$$
$$\times \exp \Biggl (2ik \Bigl [\frac{\xi a_1(\xi)}{\sqrt{s}}-\frac{\tau
a_2(\tau)}{\sqrt{s}}\Bigl]\Biggl ) \times\exp
\left[i\frac{k^2}{\sqrt{s}}(|\xi|+|\tau|)\right], \eqno(3.11)
$$
and

$$
S_{\lambda}^{(n=1)tensor}= \exp (i\lambda \kappa^2
\overline{\Pi[\nu]})
 \exp \left.\left[ 1+\frac{i\lambda^2 \kappa^4}{4} \Biggr(
\int d\eta \sum_{i=1,2}
 \Biggr(\frac{\delta\overline{\Pi[\nu]}}{\delta\nu_i(\eta)}\Biggr)^2 \Biggr )
\right]\right|_{\nu=0}. \eqno(3.12) $$

\noindent
Using Eqs $(2.27)$ and $(2.29)$, we obtain asymptotic expression for
Eqs $(3.11)$ and $(3.12)$, namely,

$$
\overline{\Pi[\nu]}=\frac{1}{(2\pi)^6s }\int d^4 k e^{-ikx}
 \int_{-\infty}^{\infty}d\xi d\tau
e^{i(k_{-}\xi-k_{+}\tau)}
a_{1}^{\mu}(\xi)a_{1}^{\nu}(\xi)D_{\mu\nu\sigma\varrho}(k)
a_{2}^{\sigma}(\tau)a_{2}^{\varrho}(\tau)$$
$$
\times \Biggl
\{1-2i\frac{k_{\bot}\triangle_{\bot}}{\sqrt{s}}[\xi\vartheta(-\xi)
+\tau\vartheta(-\tau)]+ \frac{ik^2}{\sqrt{s}(|\xi|+|\tau|)}\Biggl
\}$$

$$\approx \frac{s}{4\pi^2 }\int\frac{d^2
k_{\perp}}{k_{\perp}^2+\mu^2} \exp(ik_{\perp}x_{\perp})
+\frac{is\triangle_{\bot}}{4\pi^2 \sqrt{s}}
[x_{+}\vartheta(-x_{+})-x_{-}\vartheta(x_{-})] \int d^2
k_{\perp}\exp(ik_{\perp}x_{\perp})
\frac{k_{\perp}}{k_{\perp}^2+\mu^2}$$

$$
-  \frac{is}{8\pi^2 \sqrt{s}}(|x_{+}|+|x_{-}|)
\int\frac{d^2k_{\perp}}{k_{\perp}^2+\mu^2}
\exp(ik_{\perp}x_{\perp})=
$$
$$
=\frac{s}{2\pi }K_0(\mu|x_{\perp}|)+ \frac{s\mu}{2\pi\sqrt{s}
}\frac{\bigtriangleup_{\perp}x_{\perp}}{|x_{\perp}|}
[x_{+}\vartheta(-x_{+})-x_{-}\vartheta(x_{-})]
K_1(\mu|x_{\perp}|)$$
$$
-  \frac{is\mu^2}{4\pi\sqrt{s}}(|x_{+}|+|x_{-}|)K_0(\mu|x_{\perp}|)
 ;\eqno(3.13)$$
\noindent
Then the final expression is
$$
\frac{i\lambda^2 \kappa^4}{4}\int d\eta
\left[\Biggr(\frac{\delta\overline{\Pi[\nu]}}{\delta\nu_1(\eta)}\Biggr)^2
+\Biggr(\frac{\delta\overline{\Pi[\nu]}}{\delta\nu_2
(\eta)}\Biggr)^2 \right] \approx -\frac{i\lambda^2
\kappa^4 }{(2\pi)^8 s^2\sqrt{s}} \int d^4 k_1 d^4 k_2  \exp
[-ix(k_1+k_2)](k_1k_2)$$
$$
\times\int_{-\infty}^{\infty} d\xi_1 d\tau_1
e^{i(k_{-}^{(1)}\xi_1 -k_{+}^{(1)}\tau_1)}
\int_{-\infty}^{\infty}d\xi_2 d\tau_2
e^{ i(k_{-}^{(2)}\xi_2 -k_{+}^{(2)}\tau_2)}
\Biggl[\Phi(\xi_1,\xi_2)+\Phi(\tau_1,\tau_2)\Biggl]
$$
$$a_{1}^{\mu}(\xi_1)a_{1}^{\nu}(\xi_1)D_{\mu\nu\sigma\varrho}(k_1)
a_{2}^{\sigma}(\tau_1)a_{2}^{\varrho}
(\tau_1)a_{1}^{\rho}(\xi_2)a_{1}^{\lambda}(\xi_2)D_{\rho\lambda\eta\omega}(k_2)
a_{2}^{\eta}(\tau_2)a_{2}^{\omega}(\tau_2)$$
$$ \times [\Phi(\xi_1,\xi_2)+\Phi(\tau_1,\tau_2)]
=\frac{i\lambda^2 \kappa^4 s^2\mu^2}{8\pi^2
\sqrt{s}}(|x_{+}|+|x_{-}|) K_1^2(\mu|x_{\perp}|). \eqno(3.14)
$$

\noindent
As in the preceding section we have assumed $|x_{\perp}|\neq 0$, which ensures that all the
integrals converge. We now substitute Eqs $(3.13)$ and $(3.14)$ into
Eq.$(3.12)$ and obtain for $ S_{\lambda}^{(n=1)tensor} $ the desired
expression,

$$
S_{\lambda}^{(n=1)tensor}\approx \exp [\frac{i\kappa^2 s
\lambda}{2\pi }K_0(\mu|x_{\perp}|)] \Biggl\{1+\frac{i\kappa^2s
\lambda\mu}{2\pi \sqrt{s}}
\frac{\triangle_{\bot}x_{\bot}}{|x_{\bot}|}
[x_{+}\vartheta(-x_{+}) -x_{-}\vartheta(x_{-})]
K_1(\mu|x_{\perp}|) $$
$$ -  \frac{\kappa^2s \lambda\mu^2}{4\pi \sqrt{s}}
(|x_{+}|+|x_{-}|)K_0(\mu|x_{\perp}|) +\frac{i\kappa^4
s^2\lambda^2\mu^2}{8\pi^2 \sqrt{s}}
(|x_{+}|+|x_{-}|)K_1^2(\mu|x_{\perp}|)\Biggl \}. \eqno(3.15)
$$

\noindent
It is important to note that in contrast to the scalar model the
corresponding correction terms in quantum gravity
increase with the energy.\\
\noindent
  Using  Eq.$(3.14)$ and the phase function of the leading
eikonal behavior followed from Eq.$(3.15)$, after integration
 $dx_{+}$,  $dx_{-}$ and
$d\lambda$ for scattering amplitude in the high-energy limit $
s\gg M_{PL}^{2} \gg t $,  we obtain the following eikonal form

$$
T(s,t)^{tensor}= -2is\int d^2x_{\perp} e^{i\Delta_{\perp}x_{\perp}}
\Bigl(e^{i\chi(|x_{\perp}|s)}- 1 \Bigl), \eqno(3.16)
$$
where the eikonal phase function $ \chi(x_{\perp}s)$ by graviton
exchange increases with energy as
$$
\chi(x_{\perp}s)= \frac{\kappa^2s}{2\pi} K_{0}(\mu|\vec x_{\bot}|),
\eqno(3.17)
$$
\noindent
and in the model with vector mesons
$ \Bigr ( L_{int}=-g{\varphi}^{\star}
i{\partial}_{\sigma}{\varphi}A^{\sigma}+ g^2 A_{\sigma}A^{\sigma}
{\varphi}^{\star}{\varphi} \Bigl) $,  the eikonal phase function

$$
\chi(x_{\perp})= \frac{g^2}{2\pi} K_{0}(\mu|\vec x_{\bot}).\eqno(3.18)
$$
It should be noted that the eikonal phases given by Eqs $(2.34)$, $(3.18)$
and $(3.17)$ correspond to a Yukawa potential  between the interacting
nucleons; according to the spin of the exchange field in the scalar case
this potential decreases with energy
$ V(s,|x_{\perp}|)=-(g^2/8\pi s)(e^{-\mu |x_{\perp}|}/|x_{\perp}|)$
and is independent of energy in the vector model\linebreak
$V(s,|x_{\perp}|)=-(g^2/4\pi)(e^{-\mu |x_{\perp}|}/|x_{\perp}|)$.
In the case of  graviton  exchange a Yukawa potential
$V(s,|x_{\perp}|)=(\kappa^2 s/2\pi)(e^{-\mu |x_{\perp}|}/|x_{\perp}|)$ increases
with energy.  Comparison of these potentials has made
it possible to draw the following conclusions:  in the model with
scalar exchange, the total cross section $\sigma_t$ decreases as $1/s$,
and only the Born term predominates in the entire eikonal equation,
the vector model leads to a total cross section $\sigma_t$ tending
to a constant value as $s \rightarrow \infty$, $t/s\rightarrow 0 $.
In both the cases,the eikonal phases are purely real and consequently
the influence of inelastic scattering is disregarded in this approximation
$\sigma^{in}=0 $. In the case of graviton exchange the Froissart limit
is violated. The similar result is also obtained in work $[6]$ with the
eikonal series for Reggeized graviton exchange.\\

We may mention that in the framework of the quasipotential
approach $[29-31]$ in quantum field theory there is a rigorous
justification of the eikonal representation on the basis of the
assumption of a smooth local quasipotential. In the determination
of non-leading terms just considered we have a singular interaction
which when radiative effects are ignored leads to a singular
quasipotential of the Yukawa type which required a special care.

\section{Conclusions}

  In the framework of functional integration using the
straight-line path approximation in quantum gravity we obtained
the first-order correction terms to the leading eikonal behavior of
the Planckian-energy scattering amplitude. We also shown that
the allowance for these terms leads to the appearance of retardation
effects, which are absent in the principal asymptotic term. It is
important to note that the singular character of the correction terms
at short distances may be lead ultimately to the appearance of
non-eikonal contributions to the scattering amplitudes.
The straight-line paths approximation used in this work corresponds to
a physical picture in which colliding high energy nucleons at the act of
interaction receive  a small recoil connected with the emission of "soft"
mesons or gravitons and retain their individuality. The calculation of
non-leading terms to leading eikonal behavior
of Planckian-energy scattering can be realized by means of the
quasipotential method which provides a consistent justification
of the eikonal representation of the scattering amplitude with
the smooth local quasipotential. This problem requires some further
study.

\section{Acknowledgements}

We are  grateful to Profs. B. M. Barbashov, V. V. Nesterenko,
V. N. Pervushin for useful discussions and Prof. G. Veneziano for
suggesting this problem and encouragement.NSH is also indebted to
Profs. Zhao-bin SU, Tao XIANG, Yuan-Zhong ZHANG for support during stay at
the Institute of Theoretical Physics - Chinese Academy of Sciences
(ITP-CAS), in Beijing. This work was supported in part by ITP-CAS,
Third World Academy of Sciences and Vietnam National Research
Programme in National Sciences.\\

\end{document}